\documentclass[aps,prb,
twocolumn,
groupedaddress,showpacs,floatfix]{revtex4}

\usepackage{epsfig}
\usepackage{graphicx}
\usepackage{amsfonts}

\def\tp{${\cal T}^{(P)}$}

\def\m{${\cal M}$}

\def\tsq{${\cal T}^{*(z)}$}
\def\tsd6{${\cal T}^{*(D_6)}$}
\def\ts2f{${\cal T}^{*(2F)}$}
\def\tsa4{${\cal T}^{*(A_4)}$}
\def\tp1r{${\cal T}^{*(p1)_r}$}
\def\tsp1{${\cal T}^{*(p1)}$}
\def\tsns{${\cal T}^{*(n)}$}
\def\tnsr{${\cal T}^{*(n_r)}$}
\def\cktsa4{${\cal C}^k_{{\cal T}^{*(A_4)}}$}
\def\cstsa4{${\cal C}^s_{{\cal T}^{*(A_4)}}$}
\def\cd6{${\cal C}_{{\cal T}^{*(D_6)}}$}

\def\ep{$\mathbb{E}_\parallel$}
\def\es{$\mathbb{E}_\perp$}

\def\qd6{$q_{D_6}$}
\def\d6{$D_6$}
\def\a4{$A_4$}

\def\z6{$\mathbb{Z}^6$}
\def\ztau{$\mathbb{Z}(\tau)$}

\def\ffo{${{\mbox{$\bigcirc$}}\!\!\!\!\!\!\:{\mbox{5}}\,}$}
\def\zfo{${{\mbox{$\bigcirc$}}\!\!\!\!\!\!\:{\mbox{2}}\,}$}

\begin{document}

\title{Bulk Termination of the Quasicrystalline
Five-Fold Surface of  Al$_{70}$Pd$_{21}$Mn$_{9}$}
\author{Z. Papadopolos\footnote{Author
for correspondence:
Phone: +49 7071 29 76378; Fax: +49 7071 29 5604;
e-mail: zorka.papadopolos@uni-tuebingen.de}
and G. Kasner}
\affiliation{Institut f\"{u}r Theoretische Physik,
Universit\"{a}t Magdeburg, PSF 4120, D-39016 Magdeburg,
Germany}
\author{J. Ledieu and E.J. Cox}
\affiliation{Surface Science Research Centre,
The University of
Liverpool, Liverpool L69 3BX, UK}
\author{N.V. Richardson and Q. Chen}
\affiliation{Department of Chemistry,
University of St. Andrews,
Fife, Scotland}
\author{R.D. Diehl}
\affiliation{Department of Physics,
Pennsylvania State University,
University Park, PA 16802, USA}
\author{T.A. Lograsso and A.R. Ross}
\affiliation{Ames Laboratory, Iowa State University,
Ames, IA 50011, USA}
\author{R. McGrath}
\affiliation{Surface Science Research Centre and
Department of Physics, The University of Liverpool,
Liverpool L69 3BX, UK}
\date{\today}

\begin{abstract}

The structure of the Al$_{70}$Pd$_{21}$Mn$_{9}$ surface
has been investigated using high resolution scanning
tunnelling microscopy (STM).
 From two large five-fold terraces on the surface in a
short decorated Fibonacci sequence, atomically 
resolved surface images have been obtained. 
One of these terraces carries a rare local
configuration in a form of a ring. The location of the
corresponding sequence of terminations in the bulk 
model \m\ of icosahedral \textit{i}-AlPdMn based 
on the three-dimensional tiling \ts2f\ of an F-phase
has been estimated using this ring configuration
and the requirement from the LEED work of 
Gierer~\textit{et al.} 
that the average atomic density of the
terminations is 0.136 atoms per \AA{}$^2$.
A termination contains two atomic plane layers 
separated by a vertical distance of 0.48\AA{}.
The position of the bulk terminations is fixed within
the layers of Bergman polytopes in the model \m:
they are $4.08$ {\AA} in the direction of the bulk
from a surface of the most dense Bergman layers.
 From the coding windows of the top planes 
in
terminations in \m\  we conclude that a Penrose (P1)
tiling is possible on almost all five-fold terraces.
The shortest edge of the tiling P1, 
is either 4.8\AA{} or 7.8\AA{}.
The experimentally derived tiling of the surface
with the ring configuration has an edge-length of 
$ 8.0 \pm 0.3 $ {\AA} and hence matches the 
minimal edge-length expected from the model.

\end{abstract}

\pacs{61.44.Br, 68.35.Bs, 68.37.Ef, 61.14.Hg}

\maketitle

\section{INTRODUCTION}
\label{sec:intr}

More then ten years ago, the discovery that
centimetre size samples of
decagonal \textit{d}-AlCuCo and
icosahedral \textit{i}-AlPdMn could be grown
opened up the possibility of surface studies of these
quasicrystals~\cite{Kortan90}.
Since then other quasicrystal samples
have been grown to similar dimensions. To date,
most surface studies have been performed on the 
five-fold surface of
\textit{i}-AlPdMn~\cite{Schaub94a,Schaub94b,Schaub95,Schaub96,Urban98,Shen99,Ledieu99a,Ledieu99b,Naumovic97,Naumovic98,Naumovic99,Gierer97,Gierer98}.
A consensus has emerged from these studies that this 
surface, after fairly standard
ultra-high-vacuum (UHV) sputtering and annealing
procedures, is itself quasicrystalline.
In this work, using a combined experimental and 
theoretical approach, we show that this surface can 
be considered to be a termination of the known bulk
structure
~\cite{Boudard92,e,KPL1,KPL2,Kasner99,Papadopolos99,KPT}.

The dynamical low-energy electron diffraction (LEED) 
analysis carried out by
Gierer~\textit{et al.}~\cite{Gierer97,Gierer98} 
indicated that the five-fold
surface of the \textit{i}-AlPdMn quasicrystal
retained the bulk 
quasicrystallinity~\cite{Gierer97,Gierer98}.
X-ray photoelectron diffraction (XPD) studies are also
consistent with a quasicrystalline surface
nature~\cite{Naumovic97,Naumovic98,Naumovic99}.
Large flat terraces may be produced, and scanning
tunnelling microscopy (STM) studies have presented
similar images of the quasicrystalline surface
\cite{Schaub94a,Schaub94b,Schaub95,Schaub96,Shen99,Ledieu99a,Ledieu99b}.
  Schaub \textit{et al.}
\cite{Schaub94a,Schaub94b,Schaub95,Schaub96} produced detailed
STM images of the terraces that reveal a dense 
distribution of
{\em dark pentagonal holes}  of edge-length
circa $4.8$ {\AA} oriented parallel to each other,
together with a more random distribution of bright 
protrusions.
They correlated measurements of structural elements
both within the
terraces and across steps on the surface. 

Later, we
demonstrated a  correspondence of these measurements 
with the
geometric model \m~\cite{Kasner99,KPT,Papadopolos99}
for atomic positions of an F-phase~\cite{RMW}. 
The model \m\ is
based on the three-dimensional icosahedral tiling
\ts2f~\cite{qKPZ1} decorated essentially by 
Bergman/Mackay
polytopes~\cite{kg,e,KPL1,KPL2}.
The observed terrace structure of the surface was 
explained in terms of
the layer structure of the bulk model. 
The dark pentagons
observed on the surface corresponded to the
Bergman polytopes~\footnote{The Bergman polytope 
is a dodecahedron with particular concave pentagonal 
faces, see Ref. \cite{e}.}
in the bulk layers.
The position of a given type
of  terrace was matched to a layer characterized
by a density of certain Bergman polytopes and their
distribution pattern.
We assumed that the surface termination respects the 
integrity of the Bergman polytopes as  clusters, 
at least in the most
dense layers, and we supposed that such a 
layer of Bergman
polytopes is exactly below the termination.
However, under these assumptions
it was  not possible to explain
the observed edge-length (circa $4.8$ {\AA}) of
the dark pentagonal holes, as this was
bigger by the factor $\tau=(\sqrt{5}+1)/2$  than the
pentagonal surfaces of
the Bergman polytope (circa $3$ {\AA})~\cite{Kasner99}. 

Later Shen 
\textit{et al.},
  using an autocorrelation analysis showed that
the surface structure is consistent with a bulk 
structure based on truncated 
pseudo-Mackay icosahedra~\cite{Shen99}
and (therefore) Bergman clusters.
A fundamental limit of those previous STM studies 
(Refs 
\cite{Schaub94a,Schaub94b,Schaub95,Schaub96,Urban98,Shen99,Ledieu99a,Ledieu99b,Naumovic97}
),
 was that the resolution of the images, while 
sub-nanometre, was not atomic. Therefore direct 
comparison with bulk models was not straightforward. 
Additionally, the presence of bright
protrusions disrupted any  attempted tiling, 
and so comparison
with tiling models was not possible.
In a previous paper, we reported an improved
sample preparation
technique. This led to a more perfect surface devoid of
protrusions (section~\ref{sec:exp}), and this in turn 
led to improved resolution
in the STM images. The better resolution, 
together with the
structural perfection, allowed us to demonstrate
that the surface structure is
consistent with a bulk
termination~\cite{Ledieu01}, using the bulk model of
Boudard \textit{et al.}~\cite{Boudard92}.

In this paper we try to find the position of 
terminations in the bulk model \m\ demanding 
(i) that the terminations are ordered in a 
decorated Fibonacci sequence (sections~\ref{subsec:M},
\ref{subsec:5Pl-in-M}) as
in Refs \cite{Kasner99,Papadopolos99,KPT}, and 
(ii) that the average density of terminations is 
0.136 atoms per \mbox{\AA{}$^2$}, as determined by
Gierer~\textit{et al.}~\cite{Gierer97,Gierer98}
(section~\ref{subsec:5planes-terminations}).
The atomically resolved images of  the surface
that allow us to map the local patterns of the STM 
images (sections \ref{sec:exp}, \ref{subsec:comp})
to the local atomic configurations in the 
terminations in \m\ (section \ref{subsec:comp}),
also prove our Ansatz from section 
\ref{subsec:5planes-terminations} 
which fixes the position of the bulk termination to be 
$4.08$ {\AA} deeper within the layer of Bergman 
polytopes than we expected in 
Refs \cite{Kasner99,Papadopolos99,KPT}.
With this new  position of the termination, 
the edge-length of the
dark pentagonal holes observed
by Schaub~\textit{et al.}~\cite{Kasner99} is
now understood (section \ref{subsec:comp}).
Moreover we conclude that the termination is highly 
dense in dark pentagonal holes that we can now 
interpret as {\em dissected} Bergman (cB) polytopes 
(section \ref{subsec:comp}).
For the imaged surface terraces the densities of 
single atomic planes in corresponding terminations 
are given and their positions w.r.t. the Bergman layers 
are discussed (section \ref{subsec:comp}).

 From the general knowledge on possible tilings 
and coverings in five-fold planes in the model 
\m\  developed in section~\ref{subsec:tilings},
we analyze in section~\ref{subsec:5planes-terminations}
the possibility of the existence of the P1 tiling 
on model terminations.
 From the predicted coding windows of the top five-fold 
q-planes in terminations in \m\ 
(section \ref{subsec:comp}) 
we conclude that the Penrose (P1)
tiling is possible on almost all five-fold terraces.
In section~\ref{subsec:P1onsurfaces} we superimpose
exact patches of the P1 tiling on STM images of two
large terraces.

\section{THEORETICAL BACKGROUND}
\label{sec:theory}

\subsection{The Geometric Model \m }
\label{subsec:M}

A geometric model \m\  for the atomic positions of
\textit{i}-AlPdMn or \textit{i}-AlCuFe~\cite{KPL1,KPL2}
has been used to interpret the STM measurements data
of Schaub~\textit{et al.}~\cite{Schaub94a,Schaub95}
on the five-fold surfaces of
\textit{i}-AlPdMn~\cite{Kasner99,Papadopolos99}.
In the model \m, the $F$-phase 
\footnote{The diffraction peaks of an $F$-phase are 
labeled by 6 (half-integer) coordinates, the points 
in ${D_6}^{rec}$.  The 6-dimensional lattice
${D_6}^{rec}$ is reciprocal to the root-lattice $D_6$.
For the points in ${D_6}^{rec}$ see Table~\ref{tab:1} 
in section~\ref{subsec:5Pl-in-M}.} (see Ref. \cite{RMW})
three-dimensional tiling \ts2f~\cite{qKPZ1} is
decorated by Bergman (and automatically Mackay)
polytopes~\cite{KPL1,KPL2}.
For details on Bergman and Mackay polytopes,
see Ref.~\cite{e}.
The geometric model is based on the
Katz-Gratias model~\cite{kg} that is explained
by Elser~\cite{e} in  a three-dimensional
``parallel space'', \ep,
the space in which the model projected from $D_6$
lattice 
\footnote{A point $ q\in D_6 $ is presented by 6 
integers $(n_1,\ldots, n_6)$, such that
$\sum_{i=1}^6 {n_i}=$~{\sf even}.
} (see Refs \cite{CS,RMW}) exists.
The atoms of \textit{i}-AlPdMn~\cite{db}
or of \textit{i}-AlCuFe~\cite{kg} can be placed on  
three translational classes of atomic positions with
respect to the $D_6$-lattice, and are denoted by
$q_{D_6}(\equiv q)$, $b$ and $a$, see Table~\ref{tab:1}
in section~\ref{subsec:5Pl-in-M} and
Ref.~\cite{Papadopolos99}.
These atomic positions in \ep\ are coded by the
corresponding ``windows'' or ``acceptance domains'' in
the three-dimensional ``perpendicular space'', \es.
Note that the six-dimensional $D_6$ lattice,
which  models
an F-phase~\cite{RMW}, acts in the six-dimensional space
that is a sum of \ep\ and \es.
These windows in \es\ are denoted by $W_q$, $W_b$ and
$W_a$, respectively. The windows of the model \m\
were constructed in Refs \cite{KPL1,KPL2,Papadopolos99}.
The tiling \ts2f\  defines the quasiperiodic structure.
More accurately, the model \m\ is supported by
$\tau$\ts2f, the tiling \ts2f \ scaled by the
factor $\tau=(\sqrt{5}+1)/2$. The quasilattice points
of $\tau$\ts2f \ are in the class of $q\in D_6$.

All points of the quasilattice which contains the 
vertices of the tiling \ts2f\ can be embedded 
in a sequence of planes orthogonal to the 
five-fold symmetry axis of an
icosahedron (``five-fold direction").
The planes orthogonal to the axis
are the ``five-fold planes".
The planes appear in a sequence and have been
classified (by particular coding regions in
the window $W_{{\cal {T}}^{*{2F}} }$) into
five types, $\pm 1$, $\pm 2$, $\pm 3$, $\pm 4$, 
$\pm 5$, see Ref. \cite{Kasner99}.
The planes of type 1 to 4 are ordered in the Fibonacci
sequence with intervals $m$ and $l$.
If the planes of type 5 are included, the sequence of
planes forms a ``decorated Fibonacci sequence" with
separations $s$ (short), $m$ (medium), $l$ (long),
where $l=\tau m=\tau^2 s$, see
Fig.~\ref{fig:sequence} in
section~\ref{subsec:comp}.
How is the decorated Fibonacci sequence defined?
Let us consider the Fibonacci
sequence of intervals $M=l$ and $L=\tau l$.
If we rename $M$ by $l$ and ``decorate'' the interval
$L$ by two points, such that $L=m \bigcup s\bigcup m$,
  the decorated Fibonacci sequence appears.
For \textit{i}-AlPdMn that has the
standard distance parallel to the five-fold direction
is \ffo\ $ = 4.56$~{\AA}, and is modeled by $\tau$\ts2f,
$s=\left(\frac{2}{\tau+2}\right)$ \ffo\ $
= 2.52$~{\AA},
$m=\tau \left(\frac{2}{\tau+2}\right)$ \ffo\ $
= 4.08$~{\AA},
$l=\tau^2\left(\frac{2}{\tau+2}\right)$ \ffo\ $
= 6.60$~{\AA}.

In the planes of type 1  a
quasiperiodic tiling \tsa4~\cite{baake}
appears~\cite{PHK,Kasner99} scaled by
a factor $\tau$.  In the planes of type 2, 3 and 4,
fragments of the same tiling of a 
plane by golden triangles appear
 (see Ref.~\cite{Kasner99},
Fig.7) with the same inflation properties as in the
tiling \tsa4~\cite{baake,PHK}.

In Refs \cite{Kasner99,KPT} the model is compared to
the ideal icosahedral monograin under the assumption
that the terraces on the surface of the material are
like the planes in the bulk, i.~e. not reconstructed.
This we will first assume and then support in this
paper.
The terraces observed by Schaub~\textit{et
al.}~\cite{Schaub94a,Schaub95} were related to the
sequence of the planes of the model \m\
described above, see also Ref.~\cite{Kasner99}.
Whereas Schaub~\textit{et al.},
after annealing at $\approx 800^{\circ}C$ observed
only Fibonacci ordered step heights $m$ and $l$ on the
surface~\cite{Schaub94a,Schaub95},
Shen~\textit{at al.}, after annealing at
$\approx 630^{\circ}C$ also detected the step height
$s$, see Ref. \cite{Shen99}.

In this paper we  study the fine structure
(local atomic configurations) within the
observed terraces and  compare it to the geometric
model \m~\cite{Kasner99}.
Whereas in  Ref. \cite{Kasner99} we succeeded in 
relating the
sequence of the terrace-like five-fold surfaces of
Schaub~\cite{Schaub94a,Schaub95}  {\em to the layers}
of the Bergman polytopes in the geometric model \m,
in this paper,
using high resolution  STM images of a five-fold
surface we will fix the
position of the planes within the layers of Bergman
polytopes.

In order to recognize and identify the fine structure
of the observed surface, we  consider certain tilings
in the five-fold planes and a covering with a set of
prototiles among which are the pentagons and pentagonal
stars \footnote{These prototiles can be uniquely
reconstructed from the point set of  atomic
positions.}.
These tilings will be locally
derived from the tiling
\tsa4.  The local derivation will be exact to
a certain stage, and thereafter, random.  The
tiling \tsa4\ scaled by the factor $\tau$ defines the
quasiperiodic structure of the planes on the surfaces
according to the model \m\ introduced
above~\cite{Kasner99}. The prototiles in the tiling
\tsa4 \ are golden triangles. The edges of the triangles
in the tiling are parallel to the two-fold symmetry
axes of an icosahedron (``two-fold directions") and are
of two lengths, \zfo \ and $\tau$\zfo.
The three-dimensional  model \m\
is supported by the tiling $\tau$\ts2f \ and
consequently in the five-fold surfaces
by $\tau$\tsa4.  Hence the edges are $\tau$\zfo \ and
$\tau^2$\zfo.  With the standard value
\zfo \ $=4.795$~\AA \ in the case of \textit{i}-AlPdMn,
$\tau$\zfo \ $=7.758$~\AA \ and
$\tau^2$\zfo \ $=12.553$~\AA.

The structure on the surface observed by STM can be
tiled uniquely {\em only} if the tiling, as an
abstract structure, is derivable from the set of
quasilattice points, and if the rules of the local
derivation are defined on relatively small distances
with respect to the area of the observed surface.

\subsection{Tilings and Coverings with
             Pentagonal Prototiles contained in
             the Tiling  \tsa4 }
\label{subsec:tilings}

As an intermediate step we
locally derive the tiling \tsq \ with pentagon,
acute rhombus and hexagon as prototiles
from the quasilattice \tsa4, as shown in
Fig.~\ref{fig:fig1}.
The tiling has an inflation factor $\tau$.
It is clear that the tiling \tsq \ can be reconstructed
from its own quasilattice points.  All edges of the
prototiles in \tsq \ are of length $\tau$\zfo.
In the geometric model \m\ the prototiles are augmented
by a factor $\tau$, so the edge-length is
$\tau^2$\zfo \ $=12.553$ {\AA}.
All prototiles of \tsq\ are the unions of golden
triangles of the previous tiling \tsa4, as shown in
Fig.~\ref{fig:fig1}. If we keep that content, the
window of the tiling is identical to the window of
\tsa4 \ (because none of the vertex (quasilattice)
points is omitted).
The coding window of the tiling \tsq, without content 
of golden triangles, is shown in Fig.~\ref{fig:fig2}.
Small fractions of the tiling
\tsq \ have been observed in the five-fold surfaces
of decagonal ($d$)-AlCuCo~\cite{EST}.

%
%
\begin{figure}[ht]
\begin{center}
\epsfxsize=80mm
\caption{
The tiling \tsq \ of the plane with the acute rhombus,
pentagon and hexagon as the prototiles.
The tiles are marked by thick lines and different
gray shadows.
The tiling \tsa4, from which \tsq\ is locally derived,
is shown in background using thin lines.
}
\label{fig:fig1}
\end{center}
\end{figure}

%
%
\begin{figure}[ht]
\begin{center}
\epsfxsize=42.5mm
\caption{ The coding window of the tiling \tsq,
           without the content of golden triangles,
           is inscribed in the decagon by thick lines,
           which is the coding window of the tiling \tsa4.
           The codings of the 9 types of vertex
           configurations in the tiling \tsa4 \ are
           marked by the numbers 1-9.}
\label{fig:fig2}
\end{center}
\end{figure}

 From the intermediate tiling \tsq \ we can locally
derive a {\em covering} of the tiling \tsa4.
This covering is by two cells in the shape of pentagons,
the smaller one, $D^a_\parallel$, of
edge-length \zfo \
and the bigger,  $D^b_\parallel$,
of edge-length $\tau$\zfo, as shown in Fig.~\ref{fig:fig3}(a).
Let us denote this covering of the tiling \tsa4 \
by \cstsa4.  Each acute rhombus from \tsq \ is
transformed into a pair of pentagons of edge-length
\zfo\ (shown in the left-hand side of
Fig.~\ref{fig:fig3}(a)),
and each hexagon is transformed into a pair of
overlapping pentagons of edge-length $\tau$\zfo\
(right-hand side of Fig.~\ref{fig:fig3}(a)).
The remainder of the tiling \tsa4 \ should be
covered by pentagons of edge-length $\tau$\zfo \
as in the tiling \tsq, see Fig.~\ref{fig:fig1}.
The above-defined covering \cstsa4 \ of the tiling
\tsa4\ is a sub-covering of the covering of
Kramer~\cite{kram1,kram2}. Kramer also covers the tiling
\tsa4\  by  two pentagons of the same size as
above. These cells are
  projected Delone cells $D^a$ and $D^b$ of the
lattice $A_4$ in \ep . 
In \ep\ they are denoted by
$D^a_\parallel$ and $D^b_\parallel$, respectively.
  Let us denote Kramer's covering by the
symbol \cktsa4. The set of pentagons in \cstsa4 \ of
edge-length $\tau$\zfo \ is identical to the set
of $D^b_{\parallel}$'s in \cktsa4.
The set of pentagons in \cstsa4 \ of edge-length \zfo,
derived from the acute rhombuses, is a {\em subset}
of the set of all $D^a_{\parallel}$'s in \cktsa4,
and {\em therefore}  the covering \cstsa4 \ of \tsa4\
is a {\em subcovering} of the covering
\cktsa4~\cite{kram1,kram2}.
Whereas the  thickness 
of the covering of
\cktsa4 \ is $C^k=3 - \tau \approx 1.382$, the thickness
of the covering of the sub-covering \cstsa4 \ is
$C^s=2\tau - 2 \approx 1.236<1.382$.
(For an explanation of the thickness
of the covering see 
Ref.~\cite{book}. 
As a reference: a thickness of the covering 
of a space by a tiling always equals 1.)
In the subcovering \cstsa4 \ only the single and double
decking (covering)~\cite{book} of the tiles by the 
covering clusters are
present. The triple decking, 
which exists in the
covering \cktsa4 \ is excluded~\cite{book} in \cstsa4.
The window of the sub-covering \cstsa4 \ of
\tsa4\ by two pentagons {\em without} the content of
golden triangles is presented in Fig.~\ref{fig:fig4}.

%
%
\begin{figure}[ht]
\begin{center}
\epsfxsize=86mm
\caption{(a) The derivation
              \tsq $\longrightarrow$\cstsa4
              is in the {\em top } part of the figure;
          (b)~\tsq $\longrightarrow$\tp1r is in the
              {\em bottom } part of the figure.}
\label{fig:fig3}
\end{center}
\end{figure}

 From the tiling \tsq \ let us keep all acute rhombuses,
and replace each hexagon by two overlapping
pentagons (as in the sub-covering \cstsa4).
This is an exact local derivation, shown in the left-hand
side of Fig.~\ref{fig:fig3}(b).  At this stage we
{\em randomly} choose one of the pentagons from each
overlapping pair, and the rest of each hexagon
unites with the neighboring acute rhombus.
In this way,  either a crown or a pentagonal
star appears to replace the rhombus, and we obtain
a partly random tiling \tp1r, see the righthand side
of the Fig.~\ref{fig:fig3}(b).
The ideal class of tilings $(P1)$ with the inflation
factor $\tau$ are described in Refs \cite{GS,Ni89}.
In Fig.~\ref{fig:fig5}(a), the window that exactly
defines the quasilattice of the tiling \tsp1 \ is
inscribed in the window of the tiling \tsa4.

%
%
\begin{figure}[ht]
\begin{center}
\epsfxsize=42.5mm
\caption{ The window of the covering \cstsa4 \,
           without the content of golden triangles,
           inscribed by the thick lines in the
           window of the tiling \tsa4. }
\label{fig:fig4}
\end{center}
\end{figure}

There is another tiling of a plane by pentagonal stars,
pentagons and obtuse rhombuses introduced by
Niizeki~\cite{Ni89}.  Let us call it
the Niizeki star-tiling and denote it by \tsns.
The inflation factor of this tiling is also $\tau$.
In Fig.~\ref{fig:fig6} we derive this tiling
from the tiling \tsq.  In Fig.~\ref{fig:fig6}(a)
(top  part of Fig.~\ref{fig:fig6})
on the left-hand side, from the set of all stars,
only the locally derivable stars are presented.
The locally derivable star appears wherever
there exists an acute rhombus neighboring one
or two hexagons, each by an edge.
Between these stars, there appear obtuse rhombuses.
In Fig.~\ref{fig:fig6}(a) on the right-hand side
the white spaces around the isolated acute
rhombuses are framed by thick lines.  Inside these
patches, there appear pairs of overlapping stars,
inscribed in one single place in the figure and
marked by an arrow.  Their overlap is exactly
the acute rhombus.  Up to the choice of one star
from each pair of overlapping stars, the local
derivation of the tiling is exact.  The exact
tiling of the plane by the stars, obtuse rhombuses and
pentagons, \tsns, is uniquely determined by its window
inscribed in the window of \tsa4,
see Fig.~\ref{fig:fig5}(b).
It is the window of Niizeki tiling.
We {\em randomly} choose a star from each overlapping
pair of  stars indicated in the bottom  part of
Fig.~\ref{fig:fig6} and obtain a  partly random
tiling \tnsr.
The only edge-length that appears in the tiling
is $\tau$\zfo \ (in the geometric model \m, it is
$\tau^2$\zfo \ $=12.553$~{\AA}).
It is also the tiling that could be eventually
seen/reconstructed from the STM images of the
surfaces orthogonal to the five-fold direction in
\textit{i}-AlPdMn and \textit{i}-AlCuFe, and in a decagonal phase.

%
%
\begin{figure}[ht]
\begin{center}
\epsfxsize=42.5mm
\epsfxsize=42.5mm
\caption{(a) The window of \tsp1 \ is
              inscribed in the window of \tsa4\ by
              thick lines.
          (b) The exact tiling of the plane by the stars,
              obtuse rhombuses and pentagons, \tsns, is
              uniquely determined by its window inscribed
              by thick lines in the window of \tsa4.
              It is the window that codes the
              Niizeki tiling \tsns.}
\label{fig:fig5}
\end{center}
\end{figure}

Both exact tilings, \tsp1 \ and \tsns, can be locally
derived from their respective quasilattice points.
In the reconstruction of tilings \tsp1 \ and \tsns \
from the respective quasilattices there appear:
(i) pairs of pentagonal sets of points centered in
each other and mutually rotated by $2\pi/10$.
The set of points of the smaller size
(the smallest pentagonal set in the tiling) is on the
neighboring distances \zfo,
the bigger, on neighboring distances $\tau^{2}$\zfo.
Each pair leads to the pentagonal star;  (ii) the
isolated pentagonal sets with neighboring distances
$\tau$\zfo \ are to be connected in pentagons.
In order to reconstruct the tiling \tsp1, it is
enough to draw the pentagons from the isolated
five-tuples of five-fold symmetrically ordered points.
In order to reconstruct the tiling \tsns \ one
draws the stars from the pairs of pentagonal sets
defined above.
One can show that in an abstract sense the tilings
\tsp1 \ and \tsns \  can be mapped one-to-one to each
other~\cite{book}. If we consider an experimental 
atomically resolved five-fold surface, and tile the
observed surface, we first have to identify the
surface by a plane in the model which we will call 
an \m-plane. Then we determine
the coding window of the plane in \es, the \m-plane
window, and we place the biggest possible window of
an exact tilings \tsp1 \ or \tsns \ in the \m-plane
window, see Figs.~\ref{fig:density} and
\ref{fig:windows}.  Following these arguments,
we will determine the edge-length of a possible tiling
of an observed surface by the prototiles of the
tiling P1 in
section~\ref{sec:geom-mod}.

%
%
\begin{figure}[ht]
\begin{center}
\epsfxsize=86mm
\caption{Local derivation: 
         \tsq $\longrightarrow$ \tnsr.
         In the text, the top part of the Figure 
         is referred to as (a) and the bottom part 
         as (b).}
\label{fig:fig6}
\end{center}
\end{figure}

\subsection{Atomic Positions in Five-Fold Planes
             of the Geometric Model \m}
\label{subsec:5Pl-in-M}

In section~\ref{subsec:tilings}
we have derived the tilings \tsq, \tsp1 \
and \tsns \ either from the ideal tiling \tsa4 \ or
from their own  corresponding
quasilattices.  We have been considering exclusively
these points $q\in D_6$ that belong to the underlying
tiling of the model, $\tau$\tsa4.
Consequently the edge-lengths in both locally derived
tilings, \tp1r \ and \tnsr \ were of
length $\tau^2$\zfo \ $=12.553$\AA.
If we also take into
account the decoration of the tiling by Bergman/Mackay
polytopes, the window of the quasilattice points
of type $q\in D_6$,
$W_{q_{D_6}}$ becomes the polytope derived in
Refs  \cite{KPL1,KPL2,Papadopolos99}.

In order to study the five-fold planes of the model \m,
we present two important general facts that we
implicitly use in all our considerations.

(i)
The reciprocal lattice of the root-lattice $D_6$ 
we denote by  ${D_6}^{rec}$. The lattice ${D_6}^{rec}$
 is also known as the weight-lattice ${D_6}^w$. 
If one icosahedrally projects ${D_6}^{rec}$ to
the parallel space, \ep / (\es), an   
icosahedral \ztau-module appears~\cite{qPKDiab,moody}. 
The module points in a plane
of a three dimensional (icosahedral)
\ztau-module
in \ep, under the *~-~map~\cite{moody}, i.~e.
$\tau \longrightarrow -1/\tau$,
are mapped  in \es\ into a plane too.
The section of this plane in \es\ through the
three-dimensional window (acceptance region of the
three-dimensional quasilattice)
defines a two-dimensional window of
the quasilattice in a corresponding plane in \ep.
The analogous statement holds true for the lines.
These are the general properties
of a $\mathbb{Z}(\lambda)$ module with quadratic
irrationality $\lambda$.
In our considerations $\lambda=\tau=(\sqrt{5}+1)/2$.
The above statement is valid for the modules with
symmetries such as icosahedral, five-fold, ten-fold,
eight-fold and twelve-fold.

(ii)
Let us consider
the four translational classes with respect to
the root-lattice $D_6$
of six-dimensional points
$\frac{1}{2}(n_1,...,n_6)\in D_6^w$,
where $n_i$ are integers.
The condition for  an atomic position
$x=\frac{1}{2}(n_1,...,n_6)$
to be in a five-fold plane in \ep\ or \es\  is a
class-function presented in Table~\ref{tab:1}.
Hence, the atomic positions in a five-fold plane
of a  ${D_6}^w$-icosahedrally
projected \ztau-module belong to the
{\em single} class, $q_{D_6}(\equiv q)$, 
$b$, $a$ or $c$.
%
%
\begin{table}[!ht]
\caption{The condition for atomic positions
$x=\frac{1}{2}(n_1,...,n_6)$
to be in a five-fold plane in \ep\ or \es\  is a
class-function.
The symbols $e$ and $o$ stand for even
and odd integers, respectively. The symbol
$n^5_{\parallel}$ / $n^5_{\perp}$ is a unit normal
to the five-fold plane in \ep / \es\  space,
respectively.
$x_{\parallel}\in$ \ep\ and $x_{\perp}\in$ \es,
where $x$ is the point in six-dimensional space,
\ep + \es. The scalar product
is given in the units [${\kappa}$],
${\kappa}=1/[ \sqrt{2}(\tau+2)]$.
}
\label{tab:1}
\begin{tabular}{c|c||c|c}
{\sf  class-criterion} & {\sf class} &
{\sf  $n^5_{\parallel}\cdot
       x_{\parallel}[{\kappa}]$} &
{\sf  $n^5_{\perp}\cdot x_{\perp} [{\kappa}]$} \\
\hline \hline
{\sf  $\frac{1}{2}(e_1,...,e_6){\rm ;} \;\;
\frac{1}{2}\sum_i e_i=\;$ even } & {\sf $q_{D_6}$} &
{\sf  $e+e\tau$ } &  {\sf $e+e\tau$ }\\ 
{\sf  $\frac{1}{2}(e_1,...,e_6){\rm ;} \;\;
\frac{1}{2}\sum_i e_i=\;$ odd}  &{\sf  $b$ } &
{\sf  $e+o\tau$} & {\sf $e+o\tau$ } \\ 
{\sf  $\frac{1}{2}(o_1,...,o_6){\rm ;} \;\;
\frac{1}{2}\sum_i o_i=\;$ odd } & {\sf  $a$} &
{\sf  $o+o\tau$} & {\sf  $o+e\tau$} \\ 
{\sf  $\frac{1}{2}(o_1,...,o_6){\rm ;} \;\;
\frac{1}{2}\sum_i o_i=\;$ even } & {\sf $c$ } &
{\sf  $o+e\tau$ } & {\sf  $o+o\tau$ }   \\
\end{tabular}
\end{table}

Using the facts (i) and (ii), in the geometric
model \m\ we code each five-fold plane containing
a class of atomic positions in \ep\  by the five-fold
dissection in \es\ of the {\em single} window $W_q$, 
$W_b$ or $W_a$, corresponding
to that class.

\subsection{Densities of Five-Fold Planes and
            Terminations of Five-Fold Surfaces
            in the Geometric Model \m}
\label{subsec:5planes-terminations}

In the work of Gierer~\textit{et al.}\cite{Gierer98}
an average density of ``terminations''
of five-fold surfaces has been determined to be
$\rho_{q+b}=\rho_{q}+\rho_{b}=0.136$ atoms
per \mbox{\AA{}}$^2$.
By the density of the termination the
authors mean the sum of the densities of two
atomic planes on a surface separated by a vertical 
distance of
0.48\mbox{\AA{}}, and consequently each ``termination''
corresponds to a pair of planes separated by this 
distance.
Let us suppose that the surface (top) planes
are of type q, then the planes 0.48 \mbox{\AA{}}
below, in the geometric model \m, are of type b.
Let us calculate $\rho_{q+b}(z_\parallel)$ in the model,
where $z_\parallel$ is along a five-fold axis orthogonal
to terraces on the surface,
and let us plot this value along the corresponding
$z_\perp$, $\rho_{q+b}(z_\perp)$. The result is shown in
Fig.~\ref{fig:density}.

%
%
\begin{figure}[ht]
\begin{center}
\caption{Density $\rho_{q+b}$ of the pairs of five-fold
          planes in bulk model \m: a q-plane and
          a b-plane, 0.48\mbox{\AA{}}
          {\em below} the q-plane.
          $\rho_{q+b}$ as a function of
          $z_{\perp}$ in units of $\tau^2$ \ffo.
          The image of $z$-axes in \es, $z_{\perp}$,
          is chosen such that $z_\parallel$
          points into opposite direction of the
          bulk.
          $\rho_{q}(z_\perp)$ is the density of a 
          q-plane,
          $\rho_{b}(z_\perp)$ is the density of a 
          b-plane
          shifted by
          $c_{\perp}^{q{\rightarrow} b}
          =[\tau^4/(\tau+2)]$\ffo,
          $\rho_{q}(z_\perp)+\rho_{b}(z_\perp)
          =\rho_{q+b}(z_\perp)$.
          In the figure the {\em old} and
          the {\em new} coding regions
          of the (decorated) Fibonacci sequence of
          planes that represent the
          surface terraces in \m\ are marked. In the new region, the
          representative plane
          of the biggest clear terrace of
          Schaub~\cite{Schaub95} is marked by
          S8 on a {\em new} position. The condition
        for appearance of the Ring-plane
          in a sequence ml(R)ml is determined and
          a representative of a Ring-plane (R)
          together with a representative of the
          following Clear-plane (C) are marked.
          Finally the region of existence for P1 tilings
          on a q-plane is denoted and particular
          minimal edges are attached to their
          coding regions, see section~\ref{sec:geom-mod}.
}
\label{fig:density}
\end{center}
\end{figure}

The function $\rho_{q+b}(z_\perp)$ has 
clear (almost flat) plateau.
The appearance of the plateau is due to the polytopal
shape of the coding windows $W_q$ and $W_b$ in
the geometric model \m \cite{KPL1,KPL2,Papadopolos99}.
In particular the window $W_q$, that defines the
surface (top) plane in a termination, differs strongly
  from the spherical shape. The plateau of the
function $\rho_{q+b}(z_\perp)$ simultaneously contains
the maxima of the function and has a value that
approximately equals the average density of
terminations determined by
Gierer~\textit{et al.}\cite{Gierer98},
which is 0.136 atoms per \mbox{\AA{}}$^2$.
It is easy to conclude that all terminations on
terraces must have  equal densities. Consequently an
interval on $z_\perp$ under the plateau
(the ``carrier'' of the plateau ) {\em codes} the
terminations and indicates the {\em new}
coding of surface
q-planes to be shifted from the {\em old} value
that we expected
in Refs~\cite{KPT,Kasner99,Papadopolos99}.
In those papers~\cite{KPT,Kasner99,Papadopolos99} we
supposed that a
{\em single} surface q-plane has to have the highest
density.
In accordance with this Ansatz, at least some of the
dense layers of
Bergman polytopes were below the surfaces.
But the dark pentagons observed by
Schaub~\cite{Schaub95},
that we put in correspondence to the
Bergman polytopes
in the layer below the surface,
were bigger by a factor $\tau$ than
the faces of Bergman polytopes~\cite{Kasner99}.
Let us shift  the
surfaces of terraces by 4.08\mbox{\AA{}} in the direction
of the bulk (-4.08\mbox{\AA{}} along $z_\parallel$) 
in ``parallel'' (observable) space,
such that the q-plane on a terrace dissects the Bergman
polytopes
of the layer and the section of each Bergman polytope
is a pentagon
of edge-length $4.8$\mbox {\AA{}}, approximately
the size
of the dark pentagons observed by Schaub~\cite{Schaub95}.
This shift  in \ep\ corresponds
  to the shift  by $[2\tau / (\tau+2)]$\ffo\
along $z_\perp$ in orthogonal space.
Indeed, the coding interval of
q-planes that forces the planes on the surface to appear
in a Fibonacci sequence
(or in a decorated Fibonacci sequence) is placed
under the plateau  of the function $\rho_{q+b}(z_\perp)$ 
by this shift, see Fig.~\ref{fig:density}.
We suppose that the terrace-like five-fold
terminations do appear in a
Fibonacci (or decorated Fibonacci) sequence such that
  the top q-planes in terminations need not be 
the most dense among the q-planes,
but the above defined ``terminations'', the pairs
of planes on a surface, have the highest densities
among all such pairs of q- and b-planes in the
geometric model \m. We check our hypothesis (Ansatz) 
on two large terraces in section~\ref{sec:geom-mod}.


\section{Five-Fold Surfaces Imaged by STM;
             Surface Preparation and STM Resolution}
\label{sec:exp}

In this
section we describe the surface preparation
we have developed to obtain large flat terraces and
low surface corrugation in STM experiments. We contrast
STM results using our optimum preparation with results
previously published by us and other
groups\cite{Ledieu99a,Schaub94a,Shen99}.

Fig.~\ref{fig:STM-hrSTM} shows data from the surface of
$i$-Al-Pd-Mn after the two different preparation
procedures.
In each case the quasicrystal samples were grown at
Ames Laboratory using the Bridgman method
\cite{Delaney97,Jenks96}.  After being cut
perpendicular to their five-fold symmetry axes in air,
the sample surfaces were prepared~\cite{Ledieu01} by
polishing.  For the first preparation, Preparation~I,
the sample was polished using 6, and 1 $\mu${m}
diamond paste for one hour.  In-vacuum preparation
consisted of a few cycles of argon ion sputtering
at 1 keV energy and a normal incidence angle followed
by annealing for periods of about 1 hour at 970~K.
The results are shown in Fig.~\ref{fig:STM-hrSTM}(left hand
panels (a) and (c)). For the second preparation,
Preparation~II, a further polish using 0.25~$\mu${m}
diamond paste was used. The surface was prepared
in-vacuum by several cycles of sputtering with 0.5~keV
Ar ions, with a sputtering angle of
20$^{\circ}$-30$^{\circ}$ relative to the surface
parallel, followed by annealing to 970~K for two hours
(in total twelve hours of annealing);
Fig.~\ref{fig:STM-hrSTM}(right hand panels (b) and (d))
show the results.

When large scale scans are compared
(Fig.~\ref{fig:STM-hrSTM}(a) and (b)), it is evident
that larger terraces are obtained using Preparation~II.
For Preparation~I, the largest terraces are of the
order of 1200~\AA{} in magnitude.  For Preparation~II
terraces of width 4000~\AA{} and length of micron
size were obtained.  Further differences between the
results of the preparation techniques are observed
when scans of smaller area are compared.
Fig.~\ref{fig:STM-hrSTM}(c) and (d) show
100~\AA{}~$\times$~100~\AA{} areas of each surface.
Clearly the surface in Fig.~\ref{fig:STM-hrSTM}(c)
is not
as well resolved as that in Fig.~\ref{fig:STM-hrSTM}(d);
the bright spots in Fig.~\ref{fig:STM-hrSTM}(c)
correspond
to protrusions of height up to $2.0$~{\AA}, while
dark spots are associated with holes of depth estimated
to be at least 1.5~{\AA}.
This STM image is comparable to those in the work of
Schaub~\textit{et
al.}~\cite{Schaub94a,Schaub94b,Schaub95,Schaub96}.
This can be contrasted with the surface shown
in Fig.~\ref{fig:STM-hrSTM}(d) where there are no large
protrusions and the surface corrugation within the
terraces is $<$~1~\AA{}.
Because the STM tip can scan the surface more closely,
features on the surface are better resolved. The
features in this image have dimensions typical of
atomic sizes (\mbox{2-3~\AA{}}).
Larger features (\mbox{4-6~\AA{}}) are also
evident and probably represent
groups of a few atoms.
The LEED patterns from each of these surfaces are
qualitatively identical, but the range of electron
beam energies over which the LEED patterns are
obtained is much larger (10-300~eV) using
Preparation~II than for Preparation~I (40-180~eV).
The LEED patterns have very sharp diffraction spots,
a low background, and show five-fold symmetry.

%
%
\begin{figure}[ht]
\begin{center}
\epsfxsize=86mm
\caption{(a) 1500~{\AA}~$\times$~1500~{\AA} STM image
              showing atomically flat terraces from a
              surface prepared using Preparation~I.
          (b) 17500~{\AA}~$\times$~17500~{\AA} STM
              image showing atomically flat terraces
              from a surface prepared using
              Preparation~II.
          (c) 100~{\AA}~$\times$~100~{\AA} STM image of
              a flat terrace that we call the
              ``Clear'', C-terrace
              from a surface prepared
              using Preparation~I (bias
              voltage 2.29~V, tip current 0.59~nA).
          (d) 100~\mbox{\AA{}}~$\times$~100~\mbox{\AA{}}
              high resolution STM image of the same
              C-terrace obtained on the
              five-fold surface using Preparation~II
              (V=\mbox{1~V}, I= \mbox{0.3~nA}).}
     \label{fig:STM-hrSTM}
\end{center}
\end{figure}

The resolution can be put on a semi-quantitative basis
by calculating the two-dimensional lateral
autocorrelation functions of the images of
Fig.~\ref{fig:STM-hrSTM}(c) and (d). These are shown in
Fig.~\ref{fig:ACofSTM-hrSTM}(a) and (b) respectively.
While the symmetry of both autocorrelation patterns
is similar, the pattern of
Fig.~\ref{fig:ACofSTM-hrSTM}(b) is considerably clearer
and the correlation maxima extend to longer distances
indicating a higher degree of quasiperiodic order.

For a more quantitative comparison a radial
distribution function (RDF) has been calculated
in both cases.
The procedure consists of dividing the
\mbox{360$^{\circ}$} around the centre of the
autocorrelation function in increments.  Along each
line corresponding to each increment, the distances
from the centre to the maxima are measured. All the
measurements are then averaged and plotted
as histograms (Fig.~\ref{fig:ACofSTM-hrSTM}(c) and (d)).
It can be seen that there is considerably more
structure in the RDF in Fig.~\ref{fig:ACofSTM-hrSTM}(d)
than in that of Fig.~\ref{fig:ACofSTM-hrSTM}(c).

%
%
\begin{figure}[ht]
\begin{center}
\epsfxsize=86mm
\caption{(a) 100~\mbox{\AA{}}$~\times$~100~\mbox{\AA{}}
              lateral autocorrelation function of the STM
              image of Fig.~\ref{fig:STM-hrSTM}(c).
          (b) 100~\mbox{\AA{}}~$\times$~100~\mbox{\AA{}}
              lateral autocorrelation function of the STM
              image of Fig.~\ref{fig:STM-hrSTM}(d).
          (c) Radial distribution function calculated
              from the autocorrelation pattern of (a).
          (d) Radial distribution function calculated
              from the autocorrelation pattern of (b).}
\label{fig:ACofSTM-hrSTM}
\end{center}
\end{figure}

In summary, surfaces prepared using Preparation~II have
a much lower surface corrugation and lead to much better
resolved STM data than those previously obtained using
Preparation~I. The main differences in these procedures
are the sputtering energy and incidence angle
(suggesting that minimizing surface damage while
removing contaminants is of importance) and the
long anneal times at high temperatures which probably
serve to restore the surface composition  to that of
the bulk quasicrystal. We interpret the protrusions as
due to material on the surface which has not yet
diffused to the step edges.
A similar observation was recently made to explain the
origin of such protrusions on $d$-Al-Ni-Co surfaces
\cite{Kishida2002}.


\section{REPRESENTATIONS OF SURFACES ON TERRACES IN 
         THE GEOMETRIC MODEL \m; 
         TILING ANALYSIS OF STM IMAGES}
\label{sec:geom-mod}

\subsection{A Five-Fold Terraces Mapped to the 
         Terminations in \m}
\label{subsec:comp}

In section~\ref{subsec:5planes-terminations}
we suggested new positions of five-fold
terminations in the geometric model \m. In this section
we search for the terminations in
the geometric model \m\ on these {\em new} positions 
that fit to atomically  resolved pictures of 
five-fold surfaces on particular
terraces imaged by STM.
In a sequence of five-fold terraces we observe
a large terrace that contains a rare local
configuration that we call the ``Ring'' (R).
This configuration helps us to
{\em orient} in the bulk model \m, i.~e. to fix
the position of the R-terrace w.r.t. the five-fold
$z$-axes.
Near the R-terrace we observe the {\em clearest}
terrace that we denote by ``C''.
A fragment of the C-terrace is shown in
Fig.~\ref{fig:STM-hrSTM}(d).

On the C-terrace local configurations of the
five-fold depressions in the shape of  dark Stars (dS)
are observed. The strongly shining pentagonal local
configurations in the form of the white Flower (wF)
and the white Star pointing upwards (wSu),
both parallel to the dS
and in the same direction make a white picture
on a dark background,
see Fig.~\ref{fig:termination-clear}(a).

%
%
\begin{figure}[!ht]
\begin{center}
\caption{(a) 100~\mbox{\AA{}}~$\times$ 100~\mbox{\AA{}}
              high resolution STM image of
              the C-terrace on a five-fold surface.
              On the C-terrace frequently repeated
              local
              configurations such as a dark Star (dS),
              a white Flower (wF)
              and a white Star pointing upwards (wSu)
              parallel to the
              dS are marked. The Bergman polytope
              below
              the terrace (Bb), above the terrace (Ba),
              and the Bergman polytope dissected by the
              terrace (cB)
              are also marked. For the scale the wSu is framed
              by a pentagon of edge-length $\tau D$,
              $D\approx 4.8$\mbox{\AA{}}.
          (b) The C-terrace from (a) corresponds to
              the C-termination in \m.
              Black points are atomic positions in the
              q-1024 plane in \m\ (No 175 on
              Fig.~\ref{fig:sequence})
              which is on the surface,
              grey points are in the b-1025 plane, 0.48\AA{}
              below the q-1024 plane.
              The local
              atomic configurations
              that may present the dS,
              the wF and the wSu
              are marked. The main constituents of
              these configurations are
              the top surface of the Bergman polytopes
              that are in the layer below the surface (Bb),
              the bottom surface of the Bergman polytopes
              that are in the layer above the surface (Ba)
              and the pentagonal section of the Bergman
              polytopes from the layer that is
              dissected by the surface (cB).
              Scale: $D=\tau d=4.8$\mbox{\AA{}}.
              From the center of cB the next atomic
              position in the bulk is 2.04 \AA{}
              below the surface. }
\label{fig:termination-clear}
\end{center}
\end{figure}

In contrast to the C-terrace the
R-terrace is not continuously (globally) clear,
i.~e. the STM images of the R-terrace taken on
different places  lead to different RDFs.
Nevertheless, we observe some local configurations on
the R-terrace that are clear,
see Fig.~\ref{fig:termination-Ring}(a).
We find the white Flower (wF) and the dark Star (dS)
identical to those on the C-terrace, see 
Fig.~\ref{fig:termination-clear}(a),
but we also see a characteristic
``Ring''-configuration (R) that is present
on none of the other observed terraces.
The terrace is therefore denoted the R-terrace.
In the R-terrace there is also a configuration which 
we call the white Star pointing downwards (wSd),
it is rotated 180$^{\circ}$ w.r.t. the wSu that we 
observe on the C-terrace.
%
%
\begin{figure}[ht]
\begin{center}
\caption{(a) 75~\mbox{\AA{}}~$\times$ 75~\mbox{\AA{}}
              STM image of the R-terrace.
              The local configurations Ring (R),
              dark Star (dS) and
              white Star pointing downwards (wSd)
              are framed by three pentagons of
              edge-lengths $\tau^3 D$, $\tau^2 D$
              and $\tau D$ respectively,
              where $D\approx 4.8$\mbox{\AA{}}.
              On a bigger STM image of the R-termination
              a full white Flower (wF)
              can be seen also.
          (b) the R-terrace from (a) corresponds to the
              R-termination in \m.
              Black points are atomic positions in
              q-1037 plane in \m\
              (No 178 on Fig.~\ref{fig:sequence})
              which is on the surface,
              grey points are in b-1038 plane, 0.48\AA{}
              below the q-1037 plane.
              The local configurations of
              atomic positions
              that may represent the dark Star (dS),
              the white Flower (wF) and the
              white star (wSd) anti-parallel to the
              dark Star (dS) are marked.
              Scale: $D=\tau d=4.8$\mbox{\AA{}}.
              In the centre of the dark Star the nearest 
              atomic position is 2.04 \mbox{\AA{}}
              below the surface.
              In the q-1037 plane there are empty
              ``streets'', $\Delta = 4.56$\mbox{\AA{}}
              broad.
}
\label{fig:termination-Ring}
\end{center}
\end{figure}

As we stated the areas of both R- and C-terrace are
large and they appear in a local upward sequence
of steps ml(R)ml(C), where
$m\approx4.08$\AA{} and $l\approx\tau m$.
On the q-planes of the geometric model \m\ we find
a rare atomic configuration that may represent
a local Ring-configuration on the STM-image of
the R-terrace,
compare Figs~\ref{fig:termination-Ring}(a) and (b).
We determine
the coding of the ring-configuration (R) in \es\
and demand that the q-plane containing the
Ring-configuration
is to be found in an upward sequence of the q-planes
ml(R)ml(C) on the new positions (shifted by 4.08\AA{},
see section~\ref{subsec:5planes-terminations}), and
{\em both} R- and C-plane are to be among the planes
from the decorated Fibonacci sequence.
 From these conditions we find the coding area in \es\
along $z_\perp$ of the R-plane to be in the
interval $z_\perp\in (0.198, 0.337)[\tau^2$\ffo]
marked on Fig.~\ref{fig:density}.
In a patch of the geometric model \m\ that spreads
along $z_\parallel$-axes in an interval of 1195 \AA{}
we find only 15 representatives of the R-plane
that fulfill all conditions mentioned above.
We choose the q-1037 plane (No 178
on Fig.~\ref{fig:sequence}), the plane that is coded
in \es\ by $z_\perp=0.323 \tau^2$\ffo,
see Fig.~\ref{fig:density}.
The corresponding C-plane is then q-1024
(No 175 on Fig.~\ref{fig:sequence})
coded by $z_\perp=0.192 \tau^2$\ffo, see
Fig.~\ref{fig:density}.
In Fig.~\ref{fig:windows} the coding windows of the
R- and the C-plane are shown.
This pair of R- and C-planes (one of 15 pairs in
the model-patch)
are taken not far from the estimated model-plane S8
for Schaub's terrace No. 8~\cite{Schaub95} on a
new position q-1128 (No 193 on Fig.~\ref{fig:sequence})
coded by $z_\perp=0.211 \tau^2$\ffo,
see Fig.~\ref{fig:density}.
%
%
\begin{figure}[ht]
\begin{center}
\caption{ In \es\ the windows of the top 
          (q-)planes in R-
           and C-terminations, $W_R$ and $W_C$,
           respectively. Over them is plotted
           (i) the window of the tiling P1 of 
           edge-length 4.8\mbox{\AA{}} (in \ep) 
           denoted by $W_{(P1)}$.
           It is the maximal window of P1, such that
           $W_{(P1)}\subset W_C$ and
           (ii) the window
           of the tiling  $\tau$(P1), of edge-length
           7.8\mbox{\AA{}} (in \ep), denoted by 
           $W_{{\tau}(P1)}$.
           It is the maximal window of P1, such that
           $W_{{\tau}P1}\subset W_R(\subset W_C)$.
           The scale for the figure is set by the 
           decagon $W_{{\cal{T}}^{*(A_4)}}$, 
           which is the window  of the tiling \tsa4\ 
           with edges $d=\tau^{-1}D$ and 
           $D=4.8$\mbox{\AA{}} (in \ep) .
           For the biggest possible window
           of P1 in $W_{ {\cal T}^{*(A_4)} }$
           see Fig~\ref{fig:fig5}(a)
           in section~\ref{subsec:tilings}.
}
\label{fig:windows}
\end{center}
\end{figure}
%
%
\begin{figure}[ht]
\begin{center}
\caption{ A decorated Fibonacci sequence
           ($s=2.52$\AA{}, $m=4.08$\AA{}, $l=6.60$\AA{})
           of the q-planes
           along the $z_\parallel$ (five-fold axes)
           of type $\pm 1,\pm 2,\pm 3,\pm 4,\pm 5$
           in \m\ on the {\em old} positions,
           see Ref. \cite{Kasner99}.
           Relative to these positions the stacked 
           layers
           of the Bergman polytopes are drawn with
           their relative densities. The representative
           planes of the large R-, C- and S8-terraces
           are marked on the {\em new} positions.
           The $-4.08$\AA{} shift
           from the {\em old} to the {\em new} positions is indicated
           by arrows.
}
\label{fig:sequence}
\end{center}
\end{figure}

%
%
\begin{figure}[ht]
\begin{center}
\caption{ Radial distributions calculated
           (top) from the autocorrelation pattern of the
                 high resolution STM image shown on
                 Fig.~\ref{fig:STM-hrSTM}(d);
        (bottom) from the autocorrelation pattern of
                 the q-plane (q-1024) of the
                 C-termination in the
                 geometric model \m \ presented in
                 Fig.~\ref{fig:termination-clear}(b)
              }
\label{fig:RDF-radial1024}
\end{center}
\end{figure}

As we have stated, in contrast to the
R-terrace,  the C-terrace is uniformly clear and it has a unique
radial distribution function (RDF).
Fig.~\ref{fig:RDF-radial1024} (top) corresponds to the
RDF calculated from the high
resolution STM image of Fig.~\ref{fig:STM-hrSTM}(d)
(identical to the RDF shown in
Fig.~\ref{fig:ACofSTM-hrSTM}(d)).
Maxima are found at 7.3, 12.1, 19.4, 24.2, 31.1, and
\mbox{38.0~\AA{}} ($\pm~0.3$~{\AA}).  The radial
distribution function calculated from
the C-plane q-1024
(No 175) of the geometric model \m \  (shown in
Fig.~\ref{fig:RDF-radial1024}(bottom)), is very similar,
the main differences being the presence of a double peak
at 15~\AA{}, and some extra structure
at higher distances.
The correspondence with the
largest intensity peaks is however very good.

To the C- and the R-planes of type q, there correspond
C- and R-terminations, which are pairs of q- and b-planes
at the surface separated by a vertical distance of 0.48\AA{}.
All local patterns observed on the C-terrace
and the R-terrace
can be mapped to the model-terminations,
the C(lear)-termination and the R(ing)-termination, 
respectively,
see Figs~\ref{fig:termination-clear}(b) and
\ref{fig:termination-Ring}(b).
These patterns are mapped to the local atomic
configurations
on model-terminations that contain groups of atoms
in the shape of pentagons
related to Bergman polytopes that
(i) either {\em dissect} by the termination (cB)
(some of them are in the central parts of dS); or
(ii) are {\em below} the termination (Bb)
(some of them are in dS, wF, R, wSd); or
(iii) to the Bergman polytopes that are {\em above}
the termination (Ba) (in wSu). For this see
Figs~\ref{fig:termination-clear}(b),
\ref{fig:termination-Ring}(b) and \ref{fig:sequence}.


\subsection{The Tiling P1 on Five-Fold Surfaces}
\label{subsec:P1onsurfaces}

In order to extract information from the STM images,
  we have
employed a tiling approach 
in Refs \cite{Ledieu99a,Ledieu01}.
In Ref.~\cite{Ledieu01} this consisted of
connecting points of high contrast on the STM image
to create pentagons. The filling-in of the image using
pentagons led to a Penrose (P1)-like tiling
  of the experimental plane (with an edge-length of 
$7.8$\mbox{\AA{}}).
Here we will reconstruct exact patches of the P1 tiling
  on the STM images of both R- and C-terraces (see
Figs~\ref{fig:P1onRingSTM} and \ref{fig:P1onClearSTM})
and on corresponding model-planes (not shown).

The coding regions of
P1 tilings with minimal edge-lengths
on q-planes are marked on Fig.~\ref{fig:density}.
The tiling P1 with edge-length 7.8\mbox{\AA{}}
is coded in the interval
$z_\perp\in (-\tau^{-1}, \tau^{-1}) [\tau^2$\ffo]
and that with edge-length 4.8\mbox{\AA{}} in the 
interval
$z_\perp\in (-\tau^{-3}, \tau^{-3}) [\tau^2$\ffo].

 From the coding of the q-planes of the
C-, S8- and R-terminations
($z_{\perp}^C=0.192\tau^2$\ffo,
$z_{\perp}^{S8}=0.211\tau^2$\ffo\ and
$z_{\perp}^R=0.323\tau^2$\ffo)
we conclude that the
q-1024 plane (No. 175 in Fig.~\ref{fig:sequence})
of the C-termination and
the q-1128 plane (No. 193 in Fig.~\ref{fig:sequence})
of the S8-termination in \m\ allow a P1
tiling of minimal edge-length 4.8\mbox{\AA{}},
and the q-1037 plane (No. 178 in Fig.~\ref{fig:sequence})
of the R-termination allows a P1 tiling of minimal
edge-length 7.8\mbox{\AA{}}.
(See also in Fig.~\ref{fig:windows}
the coding windows of P1 tilings
with edge-lengths 4.8\mbox{\AA{}} and 7.8\mbox{\AA{}}
plotted over the coding windows of the q-1024 plane 
of the C-termination and the q-1037 plane of the 
R-termination).

An exact patch of the tiling P1 can be exactly
placed on the q-plane
of a model termination as follows:
(i) Plot the window of the P1 tiling, $W_{P1}$,
of the maximal possible size such that
$W_{P1}\subseteq W_{q-pl}$, where  $W_{q-pl}$
is the coding window of the surface q-plane in
the model \m. For the biggest possible window
of P1 in $W_{ {\cal T}^{*(A_4}) }$
see Fig~\ref{fig:fig5}(a)
in section~\ref{subsec:tilings}.
(ii) Mark all atomic positions coded by the
points in the window $W_{P1}$ in \ep\ . 
This set of points
uniquely determines the P1 tiling on the model q-plane.
The procedure is evident.
In contrast, for an STM image of a terrace
we have to proceed locally. If the plane is very clear
and the window $W_{P1}$ can be tightly placed in the 
corresponding
window $W_{q-pl}$ we can reconstruct an
exact patch of the P1 tiling by trial and error.
A probable exact patch of the tiling P1 with minimal
edge-length of 7.8\mbox{\AA{}} is reconstructed on
an STM image of the R-terrace,
in Fig.~\ref{fig:P1onRingSTM}.
%
%
\begin{figure}[ht]
\begin{center}
\caption{ 75~\mbox{\AA{}}~$\times$~75~\mbox{\AA{}}
           segment of an STM image of the R-terrace
           with a superimposed exact patch
           of the P1 tiling of edge-length
           $7.8$\mbox{\AA{}}.
}
\label{fig:P1onRingSTM}
\end{center}
\end{figure}

%
%
\begin{figure}[!ht]
\begin{center}
\epsfxsize=86mm
\end{center}
\begin{center}
\caption{(a) An exact patch of the P1
              tiling superimposed on the enhanced high
              resolution STM image
              (100~\mbox{\AA{}}$~\times
              $~100~\mbox{\AA{}}) of the C-terrace.
          (b) The patch of P1 tiling
              obtained from (a)
              shown superimposed on the unenhanced
              high resolution STM image of the
              C-termination from
              Fig.\ref{fig:STM-hrSTM}(d).}
\label{fig:P1onClearSTM}
\end{center}
\end{figure}

The q-1024 plane related to the surface of the
C-termination is very dense, and although we could
theoretically place the P1 tiling of minimal
edge-length 4.8\mbox{\AA{}}
(see Fig.~\ref{fig:windows}),
we have managed
to reconstruct only an exact patch of P1 tiling
of edge-length 7.8\mbox{\AA{}} on the STM image of the 
C-terrace, see
Fig.~\ref{fig:P1onClearSTM}.
For this purpose we apply an image enhancement
technique to the data of Fig.~\ref{fig:STM-hrSTM}(d)
in order to even out experimental contrast variations
(inherent in the use of the STM technique which
measures electron charge density at the surface
rather than nuclear coordinates) and to reduce
experimental noise.  The procedure is
based on Fourier filtering and consists of taking
a fast Fourier transform of the image, and then
enhancing obvious Bragg reflections with unique
$k$-values and removing experimentally-induced
diffuse features due to noise.  This modified
frequency space representation is then Fourier
transformed to obtain the filtered image shown
in Fig.~\ref{fig:P1onClearSTM}(a).
The result of this procedure is to strongly enhance
features in the image corresponding to the
selected $k$-values.  The procedure is essentially
identical to that used by Beeli and co-workers in the
enhancement of High Resolution Transmission Electron
Microscopy (HRTEM) images \cite{Soltmann2001}.

In the enhanced image the white spots that
we interpret as the images of atomic positions
are almost as sharp as in the model plane
q-1024 from the C-termination,
see Fig~\ref{fig:termination-clear}(b).
We find a patch of exact P1 tiling
of edge-length  $7.8\pm0.2$~\mbox{\AA{}} that can be
easily superimposed on the enhanced image,
see Fig.~\ref{fig:P1onClearSTM}(a).
Fig.~\ref{fig:P1onClearSTM}(b) shows this tiling
superimposed on the unenhanced STM image.


\subsection{Densities of Five-Fold Planes and
Terminations}
\label{subsec:densities}

In Table~\ref{tab:2} we compare the densities
of the R-, C- and S8-terminations, and also
the densities of single (q- and b-) planes contained in
each termination on the  old and the  new
positions.
%
%
\begin{table}[!ht]
\caption{
Densities on {\em old} and {\em new}
  positions of the R-, S8- and C-terminations 
(shifted by $-4.08$~{\AA}). Following
Gierer~\textit{et al.}~\cite{Gierer98}
a {\em termination} contains two planes on
the surface, and in \m\ its density is
$\rho_{(q+b)}=\rho_{(q)}+\rho_{(b)}$,
see section~\ref{subsec:5planes-terminations}.
}
\label{tab:2}
\begin{tabular}{c||c|c|c||c}
{\sf termination}&{\sf R}&{\sf S8}&{\sf C}&
{\sf average}\\
\hline \hline
{\sf No(Pl$_q^{old}$) } & {\sf $177$} &
{\sf $192$} & {\sf $174$} &\\
{\sf $z_\perp[\tau^2$\ffo] } & {\sf $ -0.019$} &
{\sf $-0.131$ } & {\sf $-0.150$ }&\\
\hline
{\sf ${\rho}_{(q)}^{old}$[\AA{}$^{-2}$] }&
{\sf $0.087$}&{\sf $0.084$}&{\sf $0.082$}&\\
{\sf ${\rho}_{(b)}^{old} [${\AA}$^{-2}]$ }&
{\sf $0.026$}&{\sf $0.008$}&{\sf $0.007$}&\\
{\sf ${\rho}_{(q+b)}^{old} [${\AA}$^{-2}]$ }&
{\sf $0.113$}&{\sf $0.092$}&{\sf $0.089$}
&{\sf $0.098$}\\
\hline
{\sf No(Pl$_q^{new})$ } & {\sf $178$} &
{\sf $193$ } & {\sf $175$} &\\
{\sf $z_\perp[\tau^2$\ffo] } & {\sf $0.323$} &
{\sf $0.211$} & {\sf $0.192$} &\\
\hline
{\sf ${\rho}_{(q)}^{new} $[\AA{}$^{-2}$] }&
{\sf $0.059$}&{\sf $0.074$}&{\sf $0.076$}&\\
{\sf ${\rho}_{(b)}^{new} $[\AA{}$^{-2}$] }&
{\sf $0.076$}&{\sf $0.063$}&{\sf $0.060$}&\\
{\sf ${\rho}_{(q+b)}^{new} $[\AA{}$^{-2}$] }&
{\sf $0.135$}&{\sf $0.136$}&{\sf $0.136$}
&{\sf $0.136$}\\
\end{tabular}
\end{table}
It is evident that the densities of terminations
on the {\em new} positions give a better fit to the 
LEED result of
Gierer~\textit{et al.}~\cite{Gierer98}, an average
density of 0.136 atoms per \mbox{\AA{}}$^2$.
We see that the C- and S8-terminations contain top
q-planes that are much more dense compared to the 
top q-plane of the R-termination. The STM images of
the C- and S8-terminations show that they are 
continuously clear. Another fact is that the top 
q-plane of  both (C- and S8-) terminations are on 
similar relative positions in the bulk w.r.t. 
the layers of Bergman polytopes (compare 
Fig.~\ref{fig:sequence}):
a dense layer (B1) is  below the plane,
a middle dense layer (B$^\prime$) is dissected
by the plane and a layer of low density
(B$^{\prime\prime}$) is above the plane.


\section{CONCLUSIONS}

We have presented  two  atomically resolved,
high resolution STM images of large and flat terraces
on the five-fold Al$_{70}$Pd$_{21}$Mn$_{9}$ surface.
We have mapped these surfaces to the five-fold 
terminations in the geometric model \m\ such that 
they form a decorated Fibonacci sequence, and their 
average atomic density is in agreement with the 
LEED measurements of
Gierer~\textit{et al.}~\cite{Gierer98}.
Due to the polytopal windows of the geometric 
model \m\ {\em all} terminations turn out to have 
{\em equal} and simultaneously {\em maximal} densities.
These new terminations in \m\ are placed $4.08$~{\AA}
lower than in the work of Ref. \cite{Kasner99}.
On the present STM images the
dark pentagons appear as the dark Stars.
At the new positions
of the model-termination planes the patterns of dark
pentagonal holes
are the same as in Ref. \cite{Kasner99} but now
each dark hole is of an appropriate size.
  At the new
positions the surface terminations dissect the most
dense Bergman layers in the model \m.
The local patterns on STM images are present in
the model terminations and are related to the Bergman
layers above (if one exists), below and dissected
by the termination.  Dissected  Bergman polytopes
correspond to the dark Stars.
The edge-lengths of superimposed
exact patches of the Penrose P1 tiling on two
STM images (corresponding to pentagons of height
equal to $12 \pm 0.36$~{\AA}) are shown to be in
agreement with the bulk model  \m\ based on the tiling
$\tau$\ts2f\ \cite{Kasner99}.

\section{Addendum}
Since this manuscript was submitted for publication
another paper containing STM results has been
published, see Ref. \cite{Barbier2002}.  
We note that those
authors also conclude that the $i$-Al-Pd-Mn
surface is a termination of the bulk structure.

\section{Acknowledgements}

We acknowledge to M. Boudard for putting his model at 
our disposal; the first
investigations of our STM images and comparison
to a bulk model were done on the five-fold planes of
that model.
We also acknowledge G. Booch for releasing his
software components in Ada~\cite{booch} under a
modified GPL. The code for handling the polyhedral
windows of geometric model \m\  is based partly
on these components.
The EPSRC (Grant numbers GR/N18680 and GR/N25718),
NSF (Grant number DMR-9819977),
DFG (Grant number KA 1001/4-2) and
AAT (Grant number III273-664A188816)
are acknowledged for funding.
We also acknowledge the University of St. Andrews for
financial support.


\begin{thebibliography}{10}

\bibitem{Kortan90} A.R.~Kortan, R.S.~Becker, F.A.~Thiel,
                     and H.S.~Chen, Phys. Rev. Lett. {\bf64},
                     200 (1990).

\bibitem{Schaub94a} T. M. Schaub, D. E. B\"{u}rgler,
                      H.-J. G\"{u}ntherodt, and J.-B. Suck,
                      Phys. Rev. Lett. {\bf 73},  1255  (1994).

\bibitem{Schaub94b} T. M. Schaub, D. E. B\"{u}rgler, C. Schmidt,
                      H.-J. G\"{u}ntherodt, and J.-B. Suck,
                      Z. Phys. B {\bf 96}, 93 (1994).

\bibitem{Schaub95} T. M. Schaub, D. E. B\"{u}rgler,
                     H.-J. G\"{u}ntherodt,
                     J.-B. Suck, and M. Audier,
                     Appl. Phys. {\bf A 61}, 491 (1995).

\bibitem{Schaub96} T. M. Schaub, D. E. B\"{u}rgler, C. Schmidt,
                     and H.-J. G\"{u}ntherodt, J. Non-Cryst.
                     Solids {\bf 205/207}, 748 (1996).

\bibitem{Urban98} Ph.~Ebert, F.~Yue, and K.~Urban,
                  Phys. Rev. {\bf B57}, 2821 (1998).


\bibitem{Shen99}   Z. Shen, C. R. Stoldt, C.J. Jenks,
                     T. A. Lograsso, and P. A. Thiel,
                     Phys. Rev. B {\bf 60}, 14688 (1999).

\bibitem{Ledieu99a} J. Ledieu, A. Munz, T. Parker,
                     R. McGrath,
                     R. D. Diehl, D. W. Delaney, and
                     T. A. Lograsso,
                     Surface Science {\bf 433/435},
                     665 (1999).

\bibitem{Ledieu99b} J. Ledieu, A. Munz, T. Parker, R. McGrath,
                      R. D. Diehl, D. W. Delaney,
                      and T. A. Lograsso,
                      Mat. Res. Soc. Symp. Proc. {\bf 553},
                      237 (1999).

\bibitem{Naumovic97} D. Naumovi\'{c}, P. Aebi, 
                     L. Schlapbach, and C. Beeli, 
                     {\em New Horizons in Quasicrystals: 
                     Research and Applications}
                     (A.I. Goldman, D.J. Sordelet, 
                      P.A. Thiel and J.M Dubois
                      (World Scientific), Singapore, 
                       1997).

\bibitem{Naumovic98} D. Naumovi\'{c}, P. Aebi, L. Schlapbach,
                       C. Beeli, T. A. Lograsso,
                        and D. W. Delaney,
                       {\em Proceedings of the 6th International
                       on Quasicrystals (ICQ-6, Yamada Conference
                       XLVII)} (S.~Takeuchi and T.~Fujiwara
                       (World Scientific) Singapore, 1998).

\bibitem{Naumovic99} D. Naumovi\'{c}, P. Aebi, C. Beeli, and
                       L. Schlapbach, Surface Science
                       {\bf 433-435}, 302 (1999).

\bibitem{Gierer97} M. Gierer, M.A.~Van Hove, A.I. Goldman,
                     Z. Shen, S.-L. Chang, C.J. Jenks,
                     C.-M. Zhang, and P.A. Thiel,
                     Phys. Rev. Lett. {\bf 78}, 467 (1997).

\bibitem{Gierer98} M. Gierer, M.A.~Van Hove, A.I. Goldman,
                     Z. Shen, S.-L. Chang, P.J. Pinhero,
                     C.J. Jenks, J.W. Anderegg, C.-M. Zhang,
                     and P.A. Thiel,
                     Phys. Rev. B {\bf 57}, 7628 (1998).

\bibitem{Boudard92} M. Boudard, M. de~Boissieu, C. Janot,
                      G. Heger, C. Beeli,
                      H.-U. Nissen, H. Vincent, R. Ibberson,
                      M. Audier, and J. M. Dubois,
                    J. Phys.: Cond. Matter {\bf 4}, 10149 (1992).

\bibitem{e} V.\ Elser, {\em Phil. Mag.} {\bf B73}, 641 (1996).

\bibitem{KPL1} P.\ Kramer, Z.\ Papadopolos, and
                  W.\ Liebermeister,
                in {\em Proc. of 6th Int. Conf. on Quasicrystals,
                  Yamada Conference XLVII},
                  edited by S. Takeuchi and T. Fujiwara,
                  p.71 (World Scientific, Singapore, 1998).


\bibitem{KPL2} Z.\ Papadopolos, P.\ Kramer, and
                  W.\ Liebermeister,
                  in {\em Proc. of the Int. Conf. on Aperiodic
                  Crystals, Aperiodic 1997},
                  edited by  Marc de Boissieu,
                  Jean-Louis Verger-Gaugry, and Roland Currant,
                  p. 173 (World Scientific, Singapore, 1998).

\bibitem{Kasner99} G. Kasner, Z. Papadopolos, P. Kramer,
                     and D. E. B\"{u}rgler,
                     Phys. Rev. B {\bf 60}, 3899 (1999).

\bibitem{Papadopolos99} Z. Papadopolos, P. Kramer, G. Kasner,
                          and D. B\"{u}rgler,
                          Mat. Res. Soc. Symp. Proc.
                          {\bf 553}, 231 (1999).




\bibitem{KPT} P.~Kramer, Z.~Papadopolos, and H.~Teuscher,
               J.\ Phys.:\ Condens.\ Matter\ {\bf 11}, 2729 (1999).

\bibitem{RMW} D. S. Rokhsar, N. D. Mermin and D. C. Wright,
               Phys. Rev. {\bf B35} 5487 (1987).

\bibitem{qKPZ1} P.\ Kramer, Z.\ Papadopolos, and D.\ Zeidler,
                 in  {\em Symmetries
                 in Science V: Algebraic structures, their
                 representations, realizations and physical
                 applications},
                 edited by B.\ Gruber and L. C. Biedenharn,
                 p. 395 (Plenum, New York, 1991).

\bibitem{kg} A.\  Katz,  D.\  Gratias,
              in {\em Proceedings of the 5th International 
Conference on Quasicrystals},
              edited by C.\ Janot and R.\  Mosseri
              p. 164 (World Scientific, Singapore, 1995).


\bibitem{CS} J.H. Conway and N.J.A. Sloane,
               Sphere Packings, Lattices and Groups
               (Springer, New York 1988)

\bibitem{db} M.\  de\  Boissieu, P.\ Stephens, M.\  Boudard,
                C.\  Janot, D.\ L.\  Chapman, and M.\  Audier,
          {\em J. Phys.: Condens. Matter} {\bf6}, 10725 (1994).


\bibitem{baake}  M.~Baake, P.~Kramer, M.~Schlottmann, and
                    D.~Zeidler,
                    Int.~J.~Mod.~Phys. {\bf B4}, 2217 (1990).

\bibitem{PHK} Z.~Papadopolos, C.~Hohneker, and P.~Kramer,
                    Discrete Math. {\bf 221}, 101 (2000).

\bibitem{EST} K.~Edagawa, K.~Suzuki, and S.~Takeuchi,
                  Phys. Rev. Lett. {\bf 85}, 1674 (2000).

\bibitem{kram1} P.~Kramer, J. \ Phys. {\bf A 32}, 5781 (1999).

\bibitem{kram2} P.~Kramer,
                  Mat. Sci. Eng. {\bf A 294-296} 401
                  (2000).


\bibitem{book} Z.~Papadopolos and G.~Kasner, {\em Coverings and
                 tilings derived from the class of tilings
                 \tsa4, covering of \ts2f}
                 in Coverings of Discrete Quasiperiodic Sets,
                 Theory and Applications to Quasicrystals.,
                 Edit. P.~Kramer, Z.~Papadopolos to be published
                 by Springer, Berlin 2002.




\bibitem{GS} B.~Gr\"unbaum and G.C.~Shepard,
                {\em Tilings and Patterns},
                 W.H. Freeman, San Francisco, 1987.

\bibitem{Ni89} K.~Niizeki, J. Phys. A: Math. Gen.
                 {\bf 22}, 4281 (1989).


\bibitem{qPKDiab}  Z.~Papadopolos, P.~Kramer,                    
            in {\em Proc. of Aperiodic `94}, 
            ed. by  G.\ Chapuis, W.\ Paciorek 
            (World Scientific, Singapore 1995) pp. 70-6 



\bibitem{moody}  R.\ V.\ Moody in:
                   The Mathematics of Long-Range Aperiodic Order,
                   ed. R.\ V.\ Moody, p. 403
                   Kluwer (1997).



\bibitem{Delaney97} D.W. Delaney, T.E. Bloomer, and
                      T.A. Lograsso, {\em New Horizons in
                      Quasicrystals: Research and Applications}
                      (eds. A.I. Goldman, D.J. Sordelet,
                   P.A. Thiel and J.M Dubois (World Scientific),
                      Singapore, 1997).

\bibitem{Jenks96} C. Jenks, D.W. Delaney, T.E. Bloomer,
                    S.-L. Chang, T.A. Lograsso, Z. Shen,
                    C.-M. Zhang, and P.A. Thiel,
                    Appl. Surf. Sci. {\bf 103}, 485 (1996).




\bibitem{Kishida2002} M. Kishida, Y. Kamimura, R. Tamura,
                       K. Edagawa, S. Takeuchi, T. Sato,
                       Y. Yokoyama, J.Q. Guo, A.P. Tsai,
                       Phys. Rev. B \textbf{65}
                       094208-1 (2002).


\bibitem{Ledieu01} J.~Ledieu, R.~McGrath, R.D.~Diehl,
                     T.A.~Lograsso, D.W.~Delaney,
                     Z.~Papadopolos and G.~Kasner,
                     Surface Science
                     {\bf 492/3} L729 (2001).


 
\bibitem{Soltmann2001} C. Soltmann and C. Beeli,
                        Phil. Mag. Lett. \textbf{81}
                        877 (2001).
 
\bibitem{Barbier2002} L. Barbier, D. Le Floc'h,
                       Y. Calvayrac and D. Gratias,
                       Phys. Rev. Lett.
                       \textbf{88} 085506 (2002).
 
\bibitem{booch} G. Booch: Software components with
                 Ada, Structures, Tools, and Subsystems,
                 Benjamin-Cummings Publishing Company,
                 1991.
 


\end{thebibliography}
\end{document}